\documentclass[aps,showpacs,amsmath,amssymb,twocolumn]{revtex4}

\usepackage[dvips]{graphicx}
\usepackage{dcolumn}
\usepackage{bm}
\usepackage[latin1]{inputenc}

\begin{document}

%
%
\title{Compact stars with a quark core within NJL model}

\author{C. H. Lenzi$^{1,2}$, A. S. Schneider$^3$,
  C. Provid\^encia$^2$,  R. M. Marinho Jr.$^1$}
\affiliation{$^1$Departamento de F\'{i}sica, Instituto Tecnol\'{o}gico de  Aeron\'{a}utica,
Campo  Montenegro, S\~{a}o Jos\'{e} dos Campos,
SP, 12228-900, Brazil \\
$^2$Centro de F\'{\i}sica Computacional, Department of Physics, University of Coimbra, Rua Larga, Coimbra, 3004-516, Portugal \\
$3$Department of Physics, Indiana University, Swain Hall West 117, 727 East Third Street
Bloomington, Indiana 47405}
\date{\today}

\begin{abstract}

An ultraviolet cutoff dependent on the chemical potential as proposed by
Casalbuoni {\it et al} is used in the su(3) Nambu-Jona-Lasinio model. The
model is applied to the description of stellar quark matter and compact
stars. It is shown that with a new cutoff parametrization it is
possible to obtain stable hybrid stars with a quark core. A larger cutoff at
finite densities leads to a partial chiral symmetry restoration of
quark $s$ at lower densities. A direct consequence is the onset of the $s$ quark in
stellar matter at lower densities and a softening of the equation of
state.

\end{abstract}


\maketitle

\section{\label{sec1}Introduction}

Compact stars are complex systems which may contain exotic matter such as
hyperons, kaon condensation, a non-homogenous mixed quark-hadron phase or, in their core, a pure
quark phase \cite{prak97,Glendenning}.

The hadronic phase has been successfully described within a relativistic
mean-field theory with the inclusion of hyperons (for a review see \cite{Glendenning}).  The quark phase
has frequently been described by the schematic MIT bag model
\cite{bag,alcock86} or by the Nambu-Jona-Lasinio (NJL) model
\cite{njl,njlsu3}. The NJL model contains some of the basic
symmetries of QCD, namely  chiral symmetry. It has been very successful in
describing the vacuum properties of  low lying mesons and predicts at
sufficiently high densities/temperatures a phase transition to a chiral
symmetric state \cite{Hatsuda1994,buballa99,Ruivo99,buballa}.  However, it is just an effective
theory that  does not take into account quark confinement.

The authors of \cite{Schertler1999} have studied  the possible existence of
deconfined quark matter in the interior of neutron stars using the NJL model
to describe the quark phase and could show that within this model  typical
neutron stars do not possess any deconfined quark matter in their
center. It was shown that as soon as quark matter appears the star
becomes unstable and collapses into a black-hole.  It
was also pointed out that the large constituent strange quark mass
obtained with NJL  over a wide range of
densities was the cause of this behavior. In \cite{Constanca2003}
it was shown that for warm neutrino free stellar matter a small quark
core could appear at finite temperature. The reason can be traced
back to a faster reduction of the $s$ quark constituent mass at low
densities and, therefore, the onset of the $s$ quark at lower
densities. For warm stellar matter with an entropy per particle equal
or below 2,  with or without  trapped neutrinos no quark core was obtained \cite{mp04}.

Over the last decade, it has been realized that strong interacting matter at high density  
and low temperature may possess a large assortment of  phases. Different  possible patterns for color superconductivity \cite{cs} have been conjectured 
(for a review see {\it e.g.} \cite{alf08,CN04} and references therein
quoted). Very recently, a new phase of QCD, named quarkyonic phase,  characterized by chiral symmetry and 
confinement has been predicted \cite{McP07}. We will not consider these phases in the present work.

It was shown in \cite{Casalbuoni2003} that at very large densities the standard
NJL model is not able to reproduce the correct QCD behavior of the gap
parameter in the quark color flavor locked (CFL) phase.
In order to solve this problem Casalbuoni {\it et al} have
introduced a ultraviolet cutoff dependent on the baryonic chemical
potential \cite{Casalbuoni2003}. The dependence of a parameter of the
model on the chemical potential changes the thermodynamics of the
model and has to be dealt with care \cite{goren95}. Within su(2) NJL Baldo {\it et al} have investigated whether a
cutoff dependent on the chemical potential could solve the problem
of star instability with the onset of the quark phase and concluded
that this was not a solution \cite{Baldo2007}. The question that may be raised is
whether within the su(3) NJL a different behavior occurs due to the a
different behavior of the $s$ quark constituent mass with density. We
will show that a larger cutoff  at finite baryonic densities will
move the onset of the $s$ quark to smaller densities due to a faster
decrease of the $s$ quark constituent mass with density.

After a review of the standard su(3) NJL model we will
introduce in section II the parametrization of the cutoff dependence on the chemical
potential, the thermodynamic consistency of the modified model and the
$\beta$-equilibrium conditions. In section III we
discuss the star stability and the dependence of the maximum mass
configuration on the cutoff. In the last section we draw some conclusions.


\section{The modified su(3) Nambu-Jona-Lasinio model}

\subsection{Standard su(3) NJL model}

To describe quark matter phase in neutron star, we use the su(3) NJL model with scalar-pseudoscalar and 't~Hooft six fermion interaction. The Lagrangian density of NJL model is defined by \cite{Constanca2003}:
\begin{align}\label{eq1}
 {\cal L} = & \bar\psi\left(i \gamma^\mu \partial_\mu + \hat{m}_0\right)\psi + g_s \sum_{a=0}^8 \left[\left( \bar\psi \lambda^a \psi \right)^2 + \left(\bar \psi i \gamma_5 \lambda^a \psi \right)^2 \right] \nonumber \\ 
+& g_t\left\{\det\left[ \bar\psi_i\left(1+\gamma_5\right)\psi_j \right] + \det\left[ \bar\psi_i\left(1-\gamma_5\right)\psi_j \right] \right\}, 
\end{align}
where, in flavor space, $\psi = (u;d;s)$ denotes the quark fields and
the $\lambda^a$ matrices are generators of the u(3) algebra. The term
$\hat{m}_0 = \mbox{diag}({m}_{0u},{m}_{0d},{m}_{0s})$ is
the quark current mass, which explicitly breaks the chiral symmetry of
the Lagrangian, and $g_s$ and $g_t$ are coupling constants of the model and have dimensions of mass$^{-2}$ and mass$^{-5}$, respectively.

The thermodynamic potential density $\Omega$ for a given baryonic
chemical potential $\mu$, at $T = 0$, is given by
\begin{equation}\label{termpot}
      \Omega = {\cal E} - \sum_i \mu_i \rho_i
\end{equation}
where the sum is over the quark flavors ($i$ = $u$, $d$ and $s$),  $\mu_i$ and $\rho_i$ are the chemical potential and the density, respectively, for each quark flavor $i$ and 
\begin{align}\label{eq2}
{\cal E}  = & -\eta N_c \sum_i \int^{\Lambda_{0}}_{k_{fi}} \frac{d^3p}{(2\pi)^3}\frac{p^2 + m_{0i} M_i}{E_i}
-2 g_s \sum_i \langle \bar \psi \psi \rangle_i^2 \nonumber \\ \ \ & -2 g_t \langle\bar u u \rangle \langle\bar d d \rangle \langle\bar s s \rangle- {\cal E}_0,
\end{align}
is the energy density. Above, $k_{fi} = \theta(\mu_i -
M_i)\sqrt{\mu_i^2 - M_i^2}$ is the Fermi momentum of the  quark $i$,
the constants $\eta=2$ and $N_c=3$ are the spin and color
degeneracies, respectively, and the constant ${\cal E}_0$ is included
to ensure that $\Omega = 0$ in the vacuum. The $\Lambda_0$ term is a
regularization ultraviolet cutoff to avoid divergences in the medium
integrals, and it is taken as a parameter of the model. The quark condensates and densities are defined, for each $i = u,d,s$, respectively, as
\begin{equation}\label{eq3}
 \phi_i = \langle\bar \psi \psi\rangle_i = -\eta N_c
 \int^{\Lambda_0}_{k_{fi}} \frac{p^2\, dp}{2 \pi^2} \frac{M_i}{E_i},
\end{equation}
where $M_i$ is the constituent mass of the quark $i$ and $E_i = \sqrt{p^2 + M_i^2}$, and 
\begin{equation}\label{eq4}
\rho_i = \langle \psi^\dagger \psi \rangle_i = \eta N_c
\int^{k_{fi}}_0 \frac{p^2\, dp}{2\pi^2}.
\end{equation}

Minimizing the thermodynamic potential with respect to the constituent quark mass $M_i$ results in three gap equations,
\begin{equation}\label{eq5}
M_i = m_{0i} - 4 g_s \phi_i -2g_t \phi_j \phi_k,
\end{equation}
where $i=u$, $j=d$ and $k=s$ and cyclic permutations.

As shown by authors in \cite{Constanca2003,Schertler1999,buballa99} we calculate an effective dynamical bag pressure:
\begin{equation}
B_{eff} = B_0 - B, 
\label{beff}\end{equation}
 where $B$ is given by,
\begin{align}\label{eq20}
B = & \eta N_c \sum_i \int^{\Lambda_{0}}_{0} \frac{p^2\, dp}{(2\pi)^2}\left(\sqrt{p^2 + M_i^2}-\sqrt{p^2 + m_{0i}^2}\right) \nonumber \\
\ \ &-2 g_s \sum_i \langle \bar \psi \psi \rangle_i^2 -4 g_t \langle\bar u u \rangle \langle\bar d d \rangle \langle\bar s s \rangle,
\end{align}
and $B_0 = B_{\rho_u = \rho_d = \rho_s = 0}$ is a constant.

In this work we consider the following set of parameters
\cite{Kunihiro1989,Ruivo99}: $\Lambda_0 = 631.4$ MeV, $g_s\, \Lambda^2 =1.829$,
$g_t\Lambda_0^5 = -9.4$, $m_{0u}=m_{0d}=5.6$ MeV, and $m_{0s} = 135.6$ MeV. This set of parameters was chosen in order to fit the vacuum values for the pion mass, the pion decay constant, the kaon mass, the kaon decay constant and the quark condensates: $m_\pi = 139$ MeV, $f_\pi = 93.0$ MeV, $m_k = 495.7$ MeV, $f_k = 98.9$ MeV, $\phi_{vd} = \phi_{vu} = (-246.7\,\mbox{MeV})^3$, and $\phi_{vs} = (-266.9\,\mbox{MeV})^3$. 

\subsection{Chemical potential-dependent cutoff $\Lambda(\mu)$}
As proposed by R. Casalbuoni {\it et al} in \cite{Casalbuoni2003} and
M. Baldo {\it et al} in \cite{Baldo2007}, we will introduce a chemical
potential dependency in the NJL model cutoff. This dependence
implies that the  vacuum constituent quark masses $M_{vi}$ become
chemical potential dependent and the same occurs to the coupling constants, $g_s$ and $g_t$.

In order to obtain the renormalized coupling constants we consider
that the values of the quark condensates in vacuum $\phi_{vi}$ are known
properties of the model: $\phi_{vd} = \phi_{vu} =
(-246.7\mbox{ MeV})^3$, and $\phi_{vs} = (-266.9\mbox{ MeV})^3$, 
\begin{equation}\label{fix3}
 -\eta N_c \int^{\Lambda(\mu)}_{0} \frac{p^2\, dp}{2 \pi^2} \frac{M_i(\Lambda(\mu))}{E_i(\Lambda(\mu))} = \phi_{vi}.
\end{equation}
 The new constituent quark masses $M_{vi}(\Lambda(\mu))$ are solutions
 of these equations ($M_{vu}=M_{vd}$ because $\phi_{vu}=\phi_{vd}$) and the coupling
 constants are  solutions of the two gap Eqs.(\ref{eq5})
\begin{eqnarray}
M_{vu}(\Lambda(\mu)) & = m_{0u} - 4 g_s(\Lambda(\mu)) \phi_{vu} -2g_t(\Lambda(\mu)) \phi_{vd} \phi_{vs} \nonumber \\
M_{vs}(\Lambda(\mu)) & = m_{0s} - 4 g_s(\Lambda(\mu)) \phi_{vs} -2g_t(\Lambda(\mu)) \phi_{vd} \phi_{vu}. \nonumber
\end{eqnarray}
 at $\mu = 0$,  for the constituent masses $M_{vi}(\Lambda(\mu))$ which satisfy
 Eq.(\ref{fix3}).

\begin{figure}[htb]
\includegraphics[width = 0.5\textwidth]{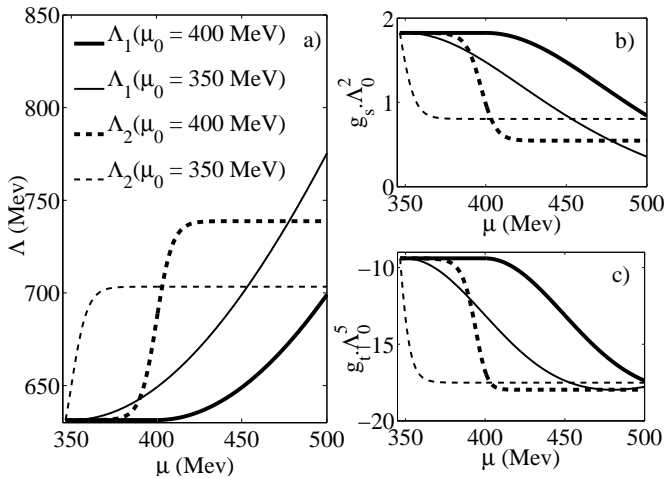}
\caption{Dependence on the chemical potential of a) the
  different parametrizations of the cutoffs discussed in the text; b)
  the $g_s$ and c) $g_t$ coupling constants.}
\label{fig1}
\end{figure}

In this paper we make two choices for the cutoff chemical potential
dependency. We use the cutoff proposed in \cite{Baldo2007},
\begin{equation}\label{eq8}
 \Lambda_1 = \left\{{ \begin{array}{lr}
\Lambda_0, \ \ \ \ \ \ \ \ \ \ \ \ & \text{if } \mu \leq \mu_0 \cr
\sqrt{9(\mu - \mu_0)^2 + \Lambda_0^2}, \ \ \  &\text{if } \mu > \mu_0
     \end{array} } \right.\
\end{equation}
where the term $\mu_0$ is the value of chemical potential above which the cutoff
becomes a function of the chemical potential. We also propose a new cutoff
\begin{equation}\label{eq9}
\Lambda_2 = \Lambda_0 + a\Lambda_0\left( \delta_0 - \frac{1}{1 + \exp{\left( \frac{\mu - \mu_0}{b}\right)}}  \right),
\end{equation}
where the constant $a$ determines the maximum value of the cutoff and
$b$ how fast the cutoff increases with density. The constant $\delta_0$ is
given by
\begin{equation}\label{eqdeita0}
 \delta_0 = \frac{1}{1 + \exp{\left( \frac{\mu_{c} - \mu_0}{b} \right)}},
\end{equation}
and ensures that $\Lambda_2(\mu_c) = \Lambda_0$. The constant $\mu_c$
is the chemical potential value for which the first  drop of quark matter
appears. P. Costa {\it et al} have calculated this parameter for each of
the quarks: $\mu_u\simeq 312$ MeV $\mu_d=\mu_s \simeq 365$ MeV
\cite{pedro2008,pedro2004} for the parameters chosen for this paper. We
assume $\mu_c = 347$ MeV, approximately equal to $(\mu_d +
\mu_u + \mu_s)/3$. In order to keep the vacuum properties the chemical
potential dependence is introduced only for $\mu\ge\mu_c$. The parameter $\mu_0$ determines the chemical potential
range where the fastest increase of the cutoff occurs.

$\Lambda_1$ is one of the choices considered in \cite{Baldo2007} which was
adjusted in order to keep the curve proposed in \cite{Casalbuoni2003} with $\Lambda_0 \simeq 580$ MeV and $\mu_0 = 400$ MeV on a range of chemical potential between $(400-600)$ MeV. In present work we are not concerned with keeping the curve proposed in \cite{Casalbuoni2003}, therefore we take different values for $\mu_0$ (between $347-400$ MeV for both cutoffs) in order to verify the effect of this parameter on the results obtained with the model. The cutoff $\Lambda_2$ of Eq.(\ref{eq9}) is a particular choice where the numerical coefficients $a=0.17$ and $b=0.005$ are adjusted in order to obtain a fast increase to the cutoff on a small interval of the chemical potential and a stabilization at some value (on the next sections we show the effect of the different values of the parameter $a$ in our results). Fig. \ref{fig1}a) shows the plots of the two different cutoffs
for two different values of $\mu_0$. For the cutoff $\Lambda_2$ the parameter $\mu_0$ does not change the rate of growth of the cutoff, however it changes the region on the chemical potential range where the increase occurs. In the same figure we can see also the
behavior of the coupling constants $g_s(\Lambda(\mu))$
[Fig. \ref{fig1}b)]  and $g_t(\Lambda(\mu))$ [Fig. \ref{fig1}c)]
as a function of the chemical potential. As discussed in \cite{Baldo2007,Casalbuoni2003} the coupling constants decrease with the increase of cutoffs in both cases.

\subsection{The thermodynamic consistency}
\begin{figure}[htb]
\begin{center}
\includegraphics[width = 0.5\textwidth]{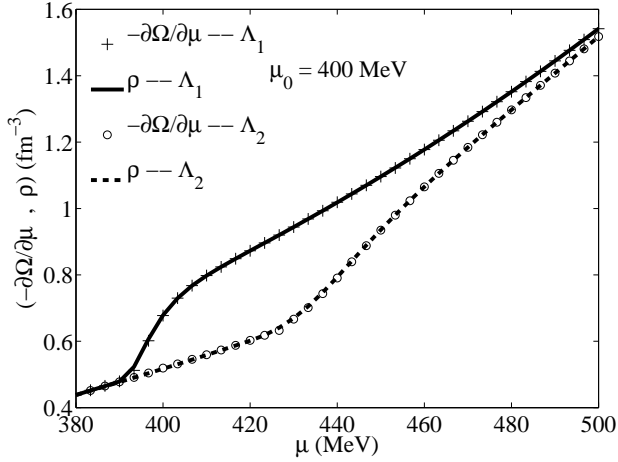}
\caption{ Plot of the thermodynamic condition,
  Eq. (\ref{eq11}), for the two parametrizations proposed
  (\ref{eq8},\ref{eq9}).}
\label{ct}\end{center}
\end{figure}

The  chemical dependent cutoff  introduced in the su(3) NJL model
gives rise to  some modifications in the thermodynamics of the system. The baryon thermodynamic potential is rewritten as
\begin{equation}\label{eq10}
 \Omega_b(k_f,\Lambda(\mu)) = {\cal E}(k_f, \Lambda(\mu)) - \sum_{i = u,d,s}\mu_i \rho_i + b(\Lambda, k_f),
\end{equation}
where the term $b(\Lambda, k_f)$ is introduced in order to maintain thermodynamical consistency
\begin{equation}\label{eq11}
\rho_i = -\frac{\partial\Omega}{\partial\mu_i}.
\end{equation}
 To calculate the the function $b$ we use the prescription of Gorenstein
and Yang \cite{goren95} and we obtain:
\begin{align}\label{eq14}
& b(\Lambda, k_f)= \frac{\eta N_c}{2\pi^2}\sum_{i=u,d,s}\int^{\Lambda}_{\Lambda_0} p^2\sqrt{p^2 + M_i^2} \ \ dp \nonumber \\ & + 2\int^{\Lambda}_{\Lambda_0} \left(\sum_{i=u,d,s}\phi_i^2 \frac{\partial g_s}{\partial\Lambda} + \phi_u \phi_d \phi_s \frac{\partial g_t}{\partial\Lambda}\right)dp,
\end{align}
so that the condition $ (\partial\Omega/\partial\Lambda)_{T,\mu}=0$.
In Fig. \ref{ct} we show that the thermodynamic condition  defined
in Eq.(\ref{eq11}) is satisfied when we include the quantity $b(\Lambda, k_f)$ in the thermodynamic potential.

\subsection{$\beta$-equilibrium condition}
In the present section we build the equation of state (EOS) of strange quark stellar matter.
We must impose both
$\beta$-{equilibrium} and electric charge neutrality
\cite{Glendenning}. We will consider cold matter,  after the  neutrinos
have diffused out and the neutrino chemical potential is zero. For $\beta$-{equilibrium} matter we  add the lepton contribution to the thermodynamic potential,
\begin{equation}\label{eq15}
\Omega(k_{fi},\Lambda(\mu)) = \Omega_b+ \Omega_l, 
\end{equation}
where $\Omega_l = {\cal E}_l(k_{fl}) - \sum_l \mu_l \rho_l$ is the
leptonic contribution taken  as that of a free Fermi gas of electrons
and muons. The electron and muon densities are
\begin{equation}\label{eq17}
 \rho_l = \frac{1}{3\pi^2}k_{fl}^3.
\end{equation}

In $\beta$-{equilibrium} the conditions of chemical equilibrium  and charge neutrality are given by
\begin{equation}\label{eq16}
 \begin{array}{c}
  \mu_s = \mu_d = \mu_u + \mu_e, \ \ \ \ \ \mu_e = \mu_\mu, \\
 \rho_e + \rho_\mu = \frac{1}{3}(2\rho_u - \rho_d - \rho_s). \\
 \end{array}
\end{equation}
\begin{figure}[htb]
\includegraphics[width = 0.5\textwidth]{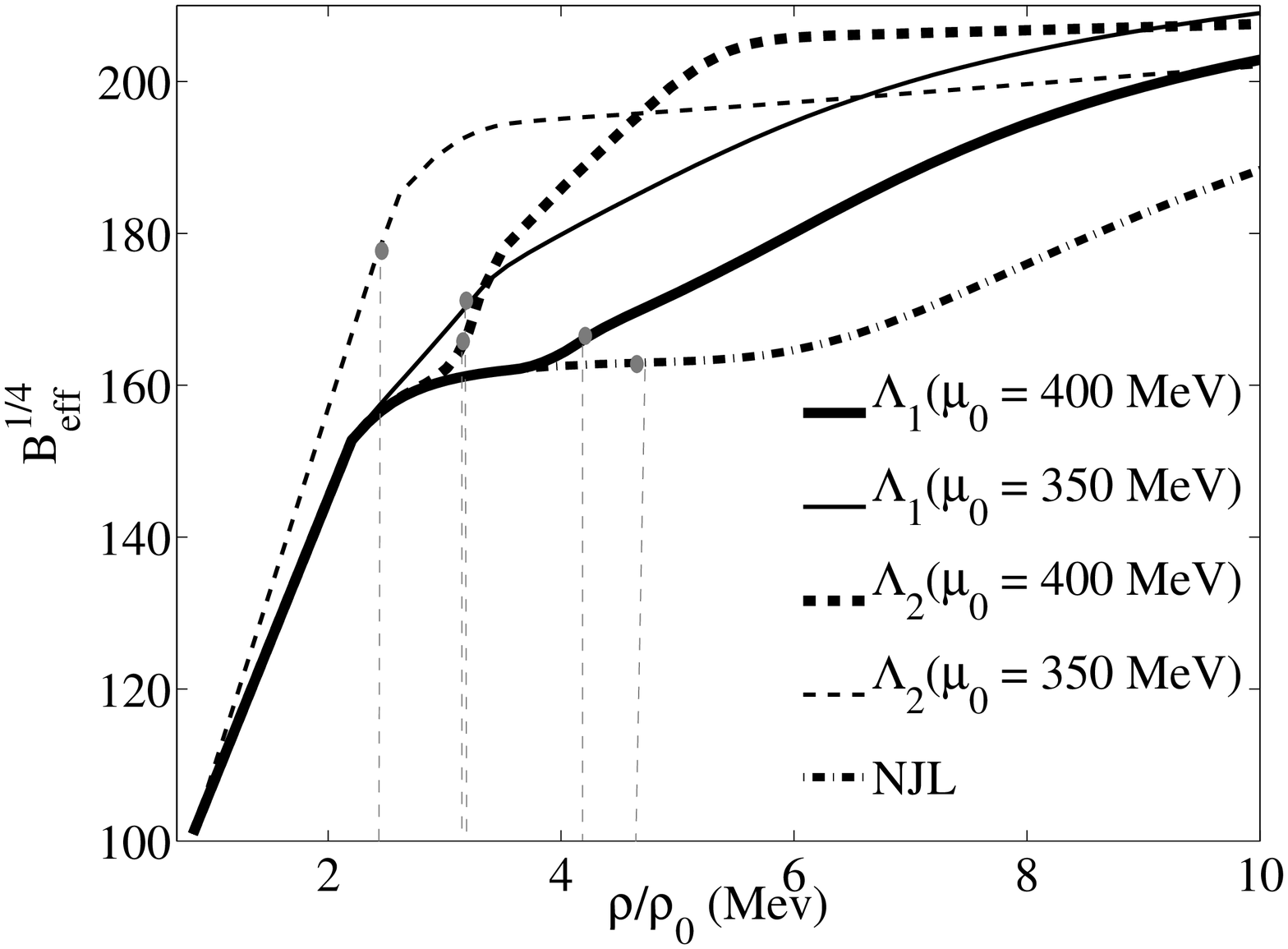}
\caption{Effective bag pressure defined in Eq. (\ref{beff})  for  different parametrizations of the
  cutoff and stellar quark matter in $\beta$-{equilibrium}.}
\label{fig2}
\end{figure}

In Fig. \ref{fig2} we plot the effective bag pressure, Eq.(\ref{eq20}),  for
each choice of the cutoff and for quark matter in $\beta$-equilibrium as a
function of the baryonic density $\rho_B = (\rho_u + \rho_d + \rho_s)/3$. As
shown in references \cite{Constanca2003,Schertler1999} there is a plateau
around $B^{1/4}=161-163$ MeV between $3\rho_0-5\rho_0$ in the standard NJL
model. The plateau is due to the partial chiral symmetry restoration of quarks
$u$ and $d$. For the new  parametrizations of the cutoff  the value
$B_{eff}^{1/4}\sim$ 162 MeV will occur at lower densities and the plateau disappears in all
cases. This effect occurs because the chemical potential dependence of the cutoff
 starts for a chemical potential 
below the partial chiral symmetry restoration for the quarks $u$ and $d$. We have marked the
onset of the strange quark on the effective bag curve with vertical lines. The effective bag
estabilizes much faster for $\Lambda=\Lambda_2$ and tends to behave in a similar way  to the MIT bag model with the decrease of the $\mu_0$. 

\begin{figure}[htb]
\includegraphics[width = 0.5\textwidth]{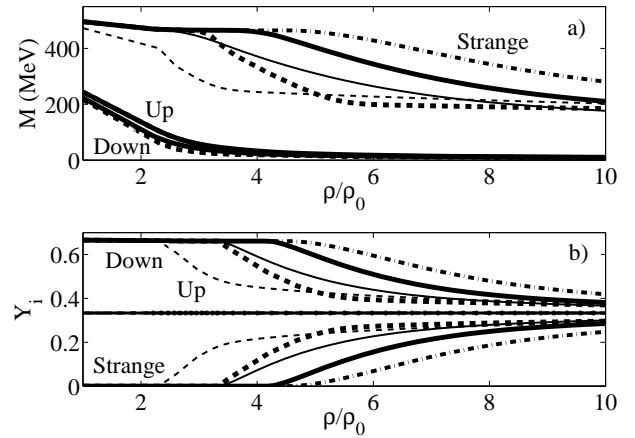}
\caption{Quark stellar matter in $\beta$-{equilibrium} with
  different choices of the cutoff: a)
fractions and b)  constituent masses of the quarks $u$, $d$, $s$ as a function of
density.}
\label{fig3}
\end{figure}

In Fig. \ref{fig3} we show the quark fractions, $Y_i =
\rho_i/(3\rho_B)$ [Fig. \ref{fig3}a)], and the constituent masses of the $u$, $d$ and $s$ quarks
in $\beta$-{equilibrium} [Fig. \ref{fig3}b)]. The effect of different choices of the cutoff on
the quark $s$ is clear: for the faster increase of  $\Lambda(\mu)$ and smaller values of $\mu_0$ the appearance of the quark $s$ occurs at lower densities and the its constituent mass $M_s$ approaches the current quark mass $m_{0s}$ at lower densities too.

\begin{figure}[htb]
\includegraphics[width = 0.5\textwidth]{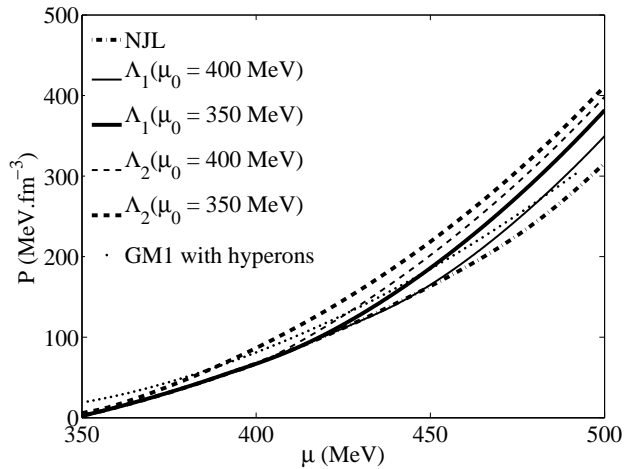}
\caption{Pressure as a function of the chemical
  potential for quark stellar matter in $\beta$-{equilibrium} with
  different choices of the cutoff and standard NJL model. The hadronic EOS,  GM1  with hyperons
\cite{Glendenning2}, is also included (dotted curve).}
\label{fig4}
\end{figure}

As can be seen from Figs. \ref{fig2} and \ref{fig3} the different slopes in the
$\mu$ dependent cutoff and the different values of the chemical potential $\mu_0$ change the
behavior of the model, namely the constituent quark masses and baryonic
density. The EOS becomes softer with these modifications, as seen in
Fig. \ref{fig4}, where the pressure is displayed as a function of the
chemical potential for the standard NJL model and the different choices of the
cutoff. The cutoff dependence proposed in Eq. (\ref{eq9}) gives the softest EOS. We also note
that decreasing $\mu_0$  favors a deconfinement phase transition at lower densities for both
parametrizations  of the cutoff, $\Lambda_1$ and$\Lambda_2$. In the same figure we plot the EOS
of the hadronic phase (dotted curve). The crossing point between
the hadronic and the quark EOS indicates the phase transition from the hadronic phase to the quark phase using a Maxwell construction \cite{Fodor}. The EOS of the quark phase constructed with standard NJL model does not cross the EOS of the hadronic phase on the chemical potential range shown.

The inclusion of a cutoff dependent on the chemical potential  in the  NJL
model influences the deconfinement phase transition  and, consequently, the stability of the star,
as we will show in the next section.

\section{The neutron star stability}

In this section we investigate the properties of stars constructed using the
modified su(3) NJL model. The Maxwell construction \cite{Fodor} is
considered for the phase transition from the hadronic phase to the quark phase. In
this case the phase transition is identified  by the crossing point between
the hadronic and the quark
EOS in the pressure versus baryonic chemical potential plane. At lower densities (below
the transition point) an hadronic phase is favored and at higher densities
(above the transition chemical potential) quark matter is favored.

However, we  should point out that the Maxwell construction is an
approximation for which only baryon number
conservation is considered and  does not take
correctly into account the existence of two charge conserving conditions, nor
surface effects and the Coulomb field \cite{vos03,maru07}.
Instead, we could have considered a Gibbs construction \cite{Glendenning},
which takes into account the existence of two charge conserving conditions. However,
a complete treatment of the mixed phase  
requires the knowledge of the surface tension between the two phases which is not well
established and may have a value between 10-100 MeV/fm$^2$ \cite{maru07}. The Gibbs construction
gives results close to the ones obtained with the lower
value of the above surface tension range, and is recovered for a zero surface
tension, while 
 it has been shown in \cite{maru07} that the Maxwell description of
the mixed phase gives a good description if the surface tension is very large.

\begin{figure}[htb]
\includegraphics[width = 0.5\textwidth]{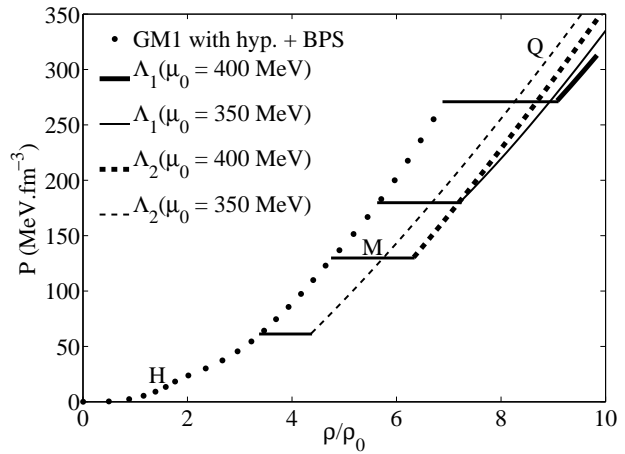}
\caption{EOS of hybrid stellar matter:  Maxwell construction for a first order
  phase transition. Pressure as a function of the baryon density for different
  parametrizations of the cutoff. The hadronic, mixed and quark phases are
  identified respectively with a H, M and Q label.}
\label{fig5}
\end{figure}

For the hadronic sector we use a EOS proposed by  Glendenning and Moszkowsky
(GM1) \cite{Glendenning2} with the inclusion of the baryonic octet. In order to fix the hyperon
coupling constants we have used one of  choices discussed in literature
\cite{Glendenning,Glendenning2}, namely we took for  all the hyperons the same coupling
constants which are a fraction $x_i$ of the meson-nucleon coupling constants, $x_\sigma=0.7$, $x_\omega=0.783=x_\rho$. For low densities (near zero density) we
use the Baym, Pethick and Sutherland (BPS) model \cite{BPS}. 
The standard and the modified su(3) NJL models are used to describe quark
matter phase. In
Fig. \ref{fig5} we plot the pressure as a function of the  baryonic density
for the complete EOS discussed above. The plateaus, identified with an M, represent the deconfinement
phase transition  as a consequence of the first order Maxwell construction. In the case of cutoff $\Lambda_1$ and lower values of $\mu_0$ the transition from hadron to quark phase occurs at lower
values of the pressure and the plateau decreases. The same situation occurs with cutoff the $\Lambda_2$.

\begin{figure}[htb]
\includegraphics[width = 0.5\textwidth]{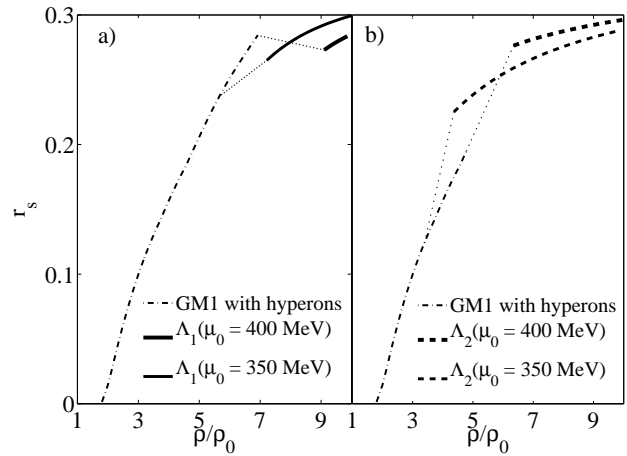}
\caption{Strangeness fraction $r_s$ for  different slopes of the cutoff for the quark phase: a) $\Lambda_1$ b) $\Lambda_2$.}
\label{strangeness}
\end{figure}

The presence of strangeness in the core and crust of the star can have an important influence in the stability of the star \cite{Alcock, Pons}. We have calculated the strangeness content of the EOS for the different parametrizations of the cutoff. In Fig. \ref{strangeness} we plot the strangeness fraction given by
\begin{equation}
r_s^{Q_S} = \frac{\rho_s}{3\rho}, \nonumber 
\end{equation}
for the quark phase and
\begin{equation}\nonumber
r_s^{Q_S} = \frac{\sum_{B}|q_s^B|\rho_B}{3\rho}, 
\end{equation}
for the hadronic phase. The term $q_s^B$ is the strange charge baryon $B$. 

The strangeness fraction is strongly modified by the different choices of the cutoff. As we can
see in Fig. \ref{strangeness}, in the case of $\Lambda_1$ with $\mu_0 = 400$ MeV, the
strangeness fraction decreases in the mixed phase and increases again in the pure quark
matter. However,   with $\mu_0 = 350$ MeV the strangeness fraction increases in the mixed phase
and continues to increase in the pure quark matter. For the case of $\Lambda_2$ the strangeness
fraction increases in both cases in the mixed phase. These different behaviors are due to the
densities at which the mixed phase occurs and the values of the constituent masses of the
strange quark for these densities. The Table \ref{t2} shows the values of the constituent
masses of strange quark on the mixed phase, the value of the densities at the onset of the
phase transition, the width of the plateau of the mixed phase and the difference of the strangeness fraction between quark and hadronic phase.
\begin{table}[htb]
\begin{center}
\caption{Constituent masses of the strange quark, densities at the onset of the phase transition, width of the plateau of the mixed phases and difference of the strangeness fraction between quark and hadronic phase for each parametrization of the cutoff.}
\begin{tabular}{c|cccc}  \hline \hline 
Cutoff                            &$M_s$  &$\rho_{QP}$ &$\Delta \rho$ & $\Delta r_s$\\
\ \                                 &(MeV)    &(fm$^{-3}$) &(fm$^{-3}$)   & ($\times 10^{-2}$)   \\  \hline  
$\Lambda_1(\mu_0 = 400\mbox{MeV})$  &229.16 & 1.00       & 0.32        &  -1.2       \\
$\Lambda_1(\mu_0 = 350\mbox{MeV})$  &225.23 & 0.82       & 0.23        &  2.9        \\
$\Lambda_2(\mu_0 = 400\mbox{MeV})$  &197.27 & 0.69       & 0.23        &  8.2        \\
$\Lambda_2(\mu_0 = 350\mbox{MeV})$  &240.61 & 0.49       & 0.14        &  10.1       \\
\hline
\end{tabular}\label{t2}
\end{center}
\end{table}

The values of the density at the onset of the phase transition  and the width  of the plateau
decrease when $\mu_0$ decreases for both parametrizations of the cutoff. On the other hand, the
discontinuity of the strangeness fraction between the two phases becomes positive and increases. The value of the
constituent mass of the s quark generally decreases if $\mu_0$ decreases except for
$\Lambda_2$ with $\mu_0 = 350$ MeV that has the biggest constituent mass due to the low density
at the phase transition density. According to references \cite{Constanca2003,Schertler1999}
these results are directly relate with the possible existence of deconfined quark matter in the
interior of neutron star as we will see later 

\begin{figure}[htb]
\includegraphics[width = 0.5\textwidth]{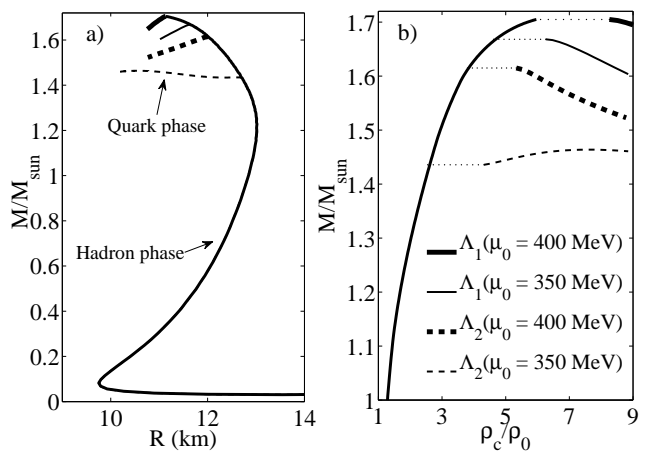}
\caption{The gravitational mass of the hybrid star is plotted as a function of
a)   the star radius and  b) the central density, for the different
parametrizations of the cutoff.}\label{massradio}
\end{figure}

We calculate the neutron star configuration for each cutoff solving the
Tolman-Oppenheimer-Volkoff (TOV) equations for a spherically symmetric and
static star\cite{Tolman,Oppenheimer}. Fig. \ref{massradio} shows the
gravitational mass of hybrid stars of the maximum mass
configuration  as a function of (a) the
radius and of (b) the central density for each
cutoff. As we can see in these plots the maximum mass is influenced by the
cutoff. In the Table \ref{t1} we show the values of the gravitational
mass, central density and the radius of the maximum mass star configuration
constructed with the EOS proposed in the present work as well as with the EOS proposed by Glendenning and Moszkowsky \cite{Glendenning,Glendenning2}.

The gravitational mass of the  hybrid stars is characterized by a cusp in the
mass versus radius plot and a plateau in the mass versus
central density graph. {The plateau is a consequence of the Maxwell construction
 and corresponds to the mixed phase between a pure hadron and a pure quark phase.
The cusp occurs at the onset of the quark phase in the interior of the star. We conclude that
for most of the models the deconfinement phase transition makes the star unstable.
} However,  for the cutoff $\Lambda_2$  with 
$\mu_0 = 350$ MeV the maximum mass configuration appears after the plateau and the cusp. In this case the
star configuration with the maximum mass has a quark phase core. We verify
that configurations with a quark phase core are possible only for the
chemical potential $\mu_0\lesssim 360$ MeV. For
values of $\mu_0\gtrsim 360$ MeV we have two possibilities: { 1) instability of the
hybrid star
with the onset of a  quark phase in the star as we can see in the case of $\mu_0 = 400$ MeV; 
2) the EOS of hadronic phase is favored for all densities if $\mu_0 > 430$ MeV.} 

\begin{figure}[htb]
\includegraphics[width = 0.5\textwidth]{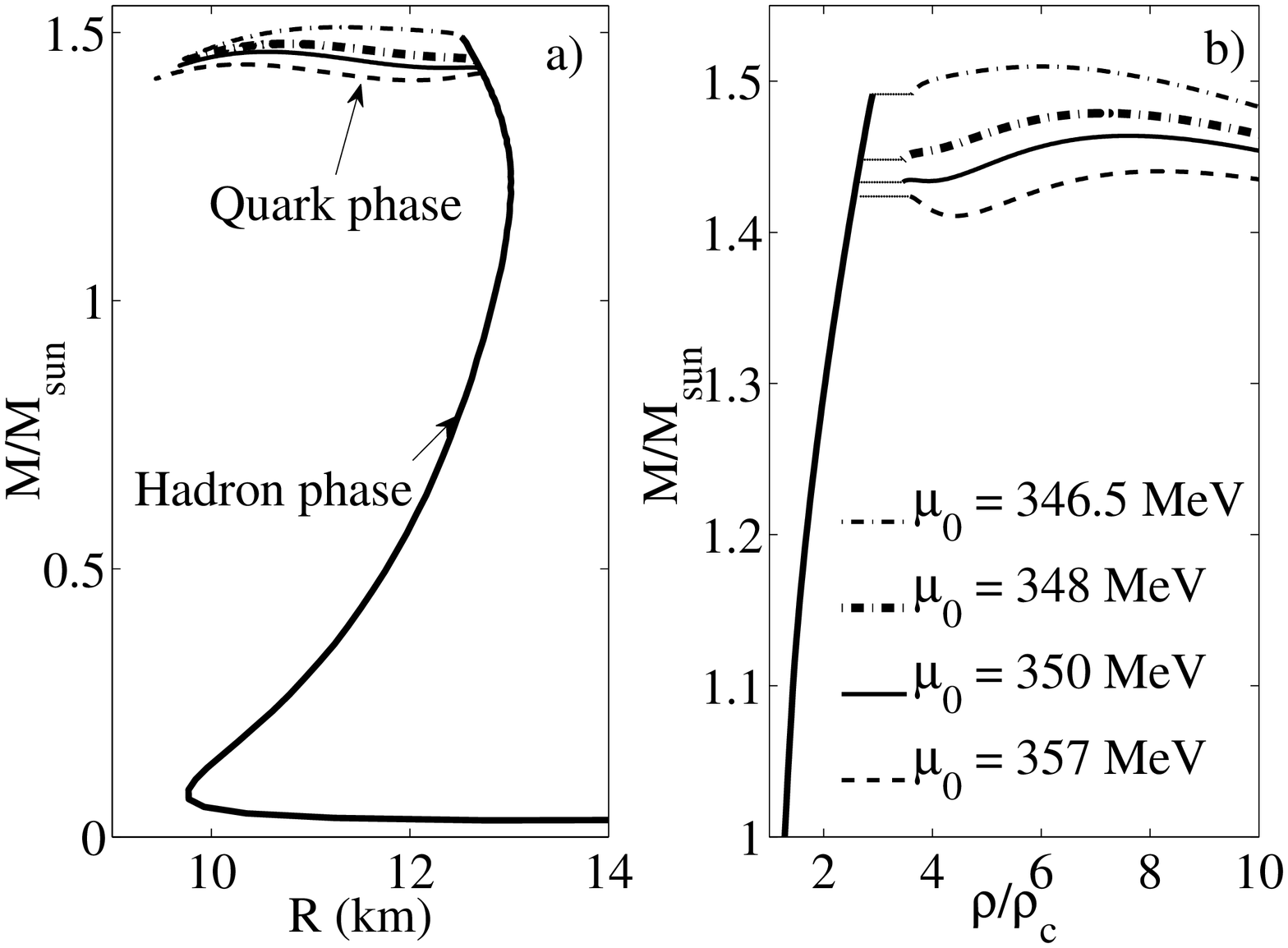}
\caption{The gravitational mass of the hybrid star is plotted as a function of
a)   the star radius and  b) the central density, for different values of $\mu_0$ for  $\Lambda_2$ with $a=0.17$.}\label{massradiomus}
\end{figure}

\begin{figure}[htb]
\includegraphics[width = 0.5\textwidth]{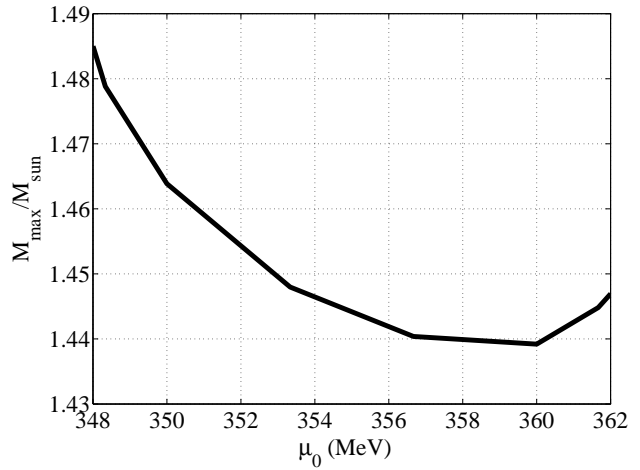}
\caption{Mass of the maximum mass star configuration as a function of the parameter $\mu_0$ for $\Lambda_2$ with $a = 0.17$.}\label{massvsmu0}
\end{figure}

\begin{figure}[htb]
\includegraphics[width = 0.5\textwidth]{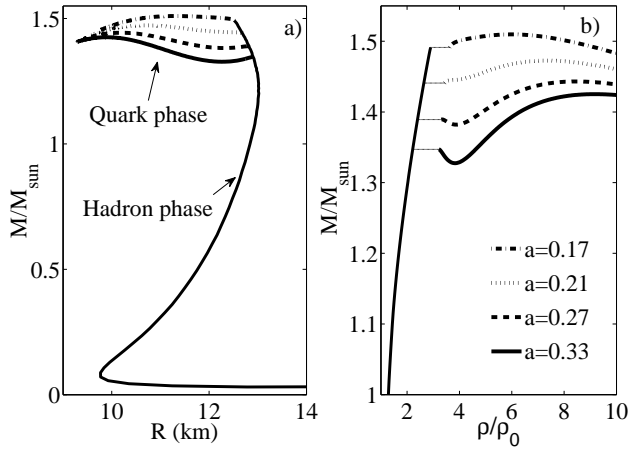}
\caption{The gravitational mass of the hybrid star is plotted as a function of
a)   the star radius and  b) the central density, for different values of $a$ for  $\Lambda_2$ with $\mu_0 = 347$ MeV.}\label{massradioas}
\end{figure}

The $\mu_0$ range with a stable quark core changes depending on the parameters used to
$\Lambda_2$. In Fig. (\ref{massradiomus}) are shown the different configurations for the
different values of $\mu_0$ in the case of $\Lambda_2$ with $a = 0.17$. The value of maximum
mass decreases if $\mu_0$ increases. We have plotted  the mass of the maximum mass
configuration as a function $\mu_0$ in Fig. \ref{massvsmu0}. The limit of stability
corresponds to the  minimum in this plot. In Fig. \ref{massradioas} the different star
configurations for different values of $a$ and the same $\mu_0=347$ MeV, are shown. We conclude
that increasing $a$ decreases the mass of the maximum mass configuration and the corresponding
radius, because the EOS becomes softer.

{It is seen from Table \ref{t1} that a smaller parameter $\mu_0$  and a
harder cutoff $\Lambda$ gives rise
to a smaller maximum mass. The occurrence of a quark core reduces a lot the maximum mass but
we can still get a reasonable value, $\sim 1.45\,\, M_\odot$, which is
consistent with the observed maximum neutron star masses, except for the
still not confirmed, 
highly massive compact stars, 
the millisecond pulsars PSR B1516 + 02B~\cite{freire1}, and PSR J1748-2021B~\cite{freire2}  with masses well above $2\, M_\odot$.}

\begin{table}[htb]
\begin{center}
\caption{Maximum gravitational mass and radius of the  hybrid stars (hs) and quark stars
  (qs) obtained with  different parametrizations of the cutoff $\Lambda$ and
  two values of the transition chemical potential $\mu_0$. In the last line the
  values for the maximum mass neutron star (ns) obtained with GM1 with hyperons
\cite{Glendenning2}. M indicates inside the mixed phase.}
\begin{tabular}{c|ccccc}  \hline \hline 
Model/Cutoff &$M_{max}$   &$R$ &$\epsilon_i$&$\epsilon_f$&$\epsilon_c$\\
\ \      &$(M_{\odot})$&(km)&(fm$^{-4}$) &(fm$^{-4}$) &(fm$^{-4}$) \\
 \hline  
$\Lambda_1(\mu_0 = 400\mbox{MeV})$  & 1.701 & 11.14 & 5.97 & 8.26 & M \\
$\Lambda_1(\mu_0 = 350\mbox{MeV})$  & 1.674 & 11.64 & 4.69 & 6.22 & M \\
$\Lambda_2(\mu_0 = 400\mbox{MeV})$  & 1.621 & 12.00 & 3.84 & 5.29 & M \\
$\Lambda_2(\mu_0 = 350\mbox{MeV})$  & 1.456 & 10.56 & 2.60 & 3.42 & 7.62 \\
\hline GM1 with hyperons            & 1.705 & 11.11 & & &  5.99 \\
\hline
\end{tabular}\label{t1}
\end{center}
\end{table}

\begin{figure}[htb]
\includegraphics[width = 0.5\textwidth]{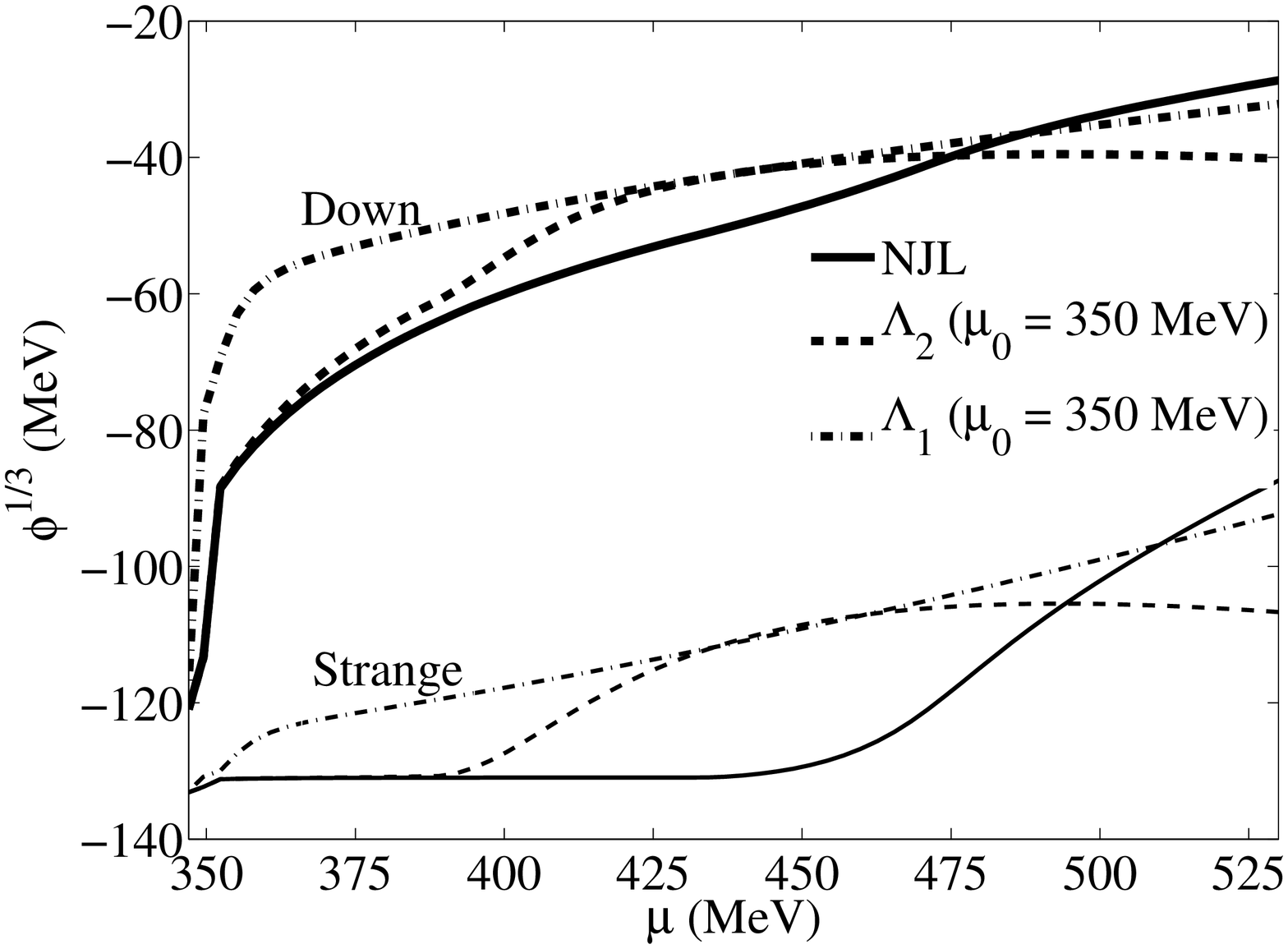}
\caption{Quark $u$ and $s$ condensates versus density for the  different cutoff parametrizations.}\label{condensate}
\end{figure}

Fig. \ref{condensate} shows the  $u$ and $s$   quark condensates as a function of the chemical
potential for both cutoff parametrizations. For the cutoff $\Lambda_1$ the module of the
condensates starts to increase for $\mu\gtrsim 475$ MeV. This is not the  case  $\Lambda_2$:
the module of the condensates decreases with density leading  the system to chiral symmetry
restauration. Therefore, the cutoff proposed in this work is physically favored to  cutoff
$\Lambda_1$. We believe the behavior of $\Lambda_1$ at high densities is due to the high value
of the cutoff $\Lambda_1$ at these densities.

\section{Summary}
We have studied the possibility of formation of stable compact stars with a
quark core within the su(3) NJL model with a chemical potential
dependent ultra-violet cutoff.  We use a su(3) NJL model parametrization which
describes the vacuum properties of low mass mesons (pions and kaons) and choose the parametrization of the
cutoff so that it increases with density. One of the consequences of increasing the  cutoff  is a faster decrease the constituent $s$ quark mass 
with density and, therefore, the onset of $s$ quark at lower densities, giving rise
to a larger pressure for the same chemical potential. The phase
transition to a deconfined quark phase occurs at smaller densities and  pressures and  the
density discontinuity at the phase transition is smaller. For cutoff $\Lambda_2$ $(a=0.17)$, stars with a quark core
are obtained for a choice of the parameter $\mu_0<360$ MeV. The maximum mass of these stars is between (1.46-1.51) $M_\odot$, and is compatible
with most of the compact star observations. However, the highly massive stars
  PSR B1516 +
02B~\cite{freire1}, and  PSR J1748-2021B~\cite{freire2}, in case they
are confirmed, would not reproduced.

According to Baldo {\it et al} the instability of compact stars
within the
NJL
model is probably due to the lack of confinement in this model
\cite{Baldo2007}, since the authors of \cite{thomas} were able to obtain 
stable stars with a quark core introducing a confining potential in the NJL 
model. The confining potential in the approach of \cite{thomas} is switched 
off at the chiral phase transition. In \cite{Baldo2007} the authors have tried
to get stable compact stars with a quark core using su(2) NJL model with a
cutoff dependent on the chemical potential and were not successful. Using the
same dependence of the chemical potential, but introducing also the strange
flavor we were also not able to obtain stable compact stars with a quark core. 
However, when we use the new cutoff proposed in this paper it is possible to get
stable compact stars with a quark core if the strange flavor is included. We believe 
the stability of core quark occurs due to the fast increase of the cutoff $\Lambda_2$ 
allowing for a chiral symmetry restoration for the $s$-quark at much lower densities 
than the ones predicted by a constant cutoff. The result is an EOS soft enough to give 
rise to a quark core stable in a hybrid star. The stabilization of the cutoff at high densities
is an important characteristic because it repares divergence  problems due to the fast increase of the cutoff.

The effect of color superconductivity was not considered in the present work
and will be the subject of a future work.

\begin{acknowledgements}
We would like to thank the fruitful discussions with Jo\~ao  da
 Provid�ncia, Pedro Costa  and Tobias Frederico. 
This work was partially supported by FEDER and Projects PTDC/FP/64707/2006 and
CERN/FP/83505/2008, and  by COMPSTAR, an ESF Research Networking Programme. CHL thanks to CAPES
by the fellowship 2071/07-0 and the international cooperation program Capes-FCT between
Brazil-Portugal.

\end{acknowledgements}



\clearpage


\appendix

\end{document}